\documentclass[epj]{svjour}

\usepackage{graphics}
\usepackage{hyperref}

\begin{document}

\title{Contribution of the non-linear term in the Balitsky-Kovchegov equation to the nuclear structure functions}
\author{Jan Cepila \and Marek Matas}                  

\institute{Faculty of Nuclear Sciences and Physical Engineering, Czech technical University in Prague, Czech Republic}

\abstract{
In this paper, we present nuclear structure functions calculated from the impact-parameter dependent solution of the Balitsky-Kovchegov equation with our recently proposed set of nuclear initial conditions. We calculate the results with and without the non-linear term in the BK equation in order to study the impact of saturation effects on the measurable structure functions and nuclear modification factor. The difference of these results rises with decreasing Bjorken $x$ and increasing scale. These predictions are of interest to the physics program at the future ep and eA colliders.
} 
\authorrunning{J. Cepila \and M. Matas}
\titlerunning{Contribution of the non-linear term in the BK equation to the nuclear structure functions}
\maketitle

\section{Introduction}
\label{intro}
With the recently approved Electron Ion Collider in the USA \cite{Accardi:2012qut} and planned LHeC \cite{AbelleiraFernandez:2012cc} at CERN, a new interest is sparked in understanding the difference between the structure function of nuclear and of proton targets. At small values of Bjorken $x$, the nuclear structure function per one nucleon is smaller than the nucleon structure function. This effect, called shadowing, can be partly explained as a result of gluon recombination due to the overlap of the gluon wave functions from the surrounding nucleons in the frame where the target is moving very fast \cite{Gribov:1984tu,Mueller:1985wy}. In this way, the gluon density in a bound nucleon is smaller than the gluon density in a free nucleon. 

This phenomenon is called saturation since at certain saturation scale the recombination processes balance gluon splitting, effectively saturating the gluon density. Quantitatively, the evolution of gluon density in this frame is described by non-linear evolution equations \cite{Gribov:1984tu,Mueller:1985wy}. Recent review of available evolution equations can be found in e.g. \cite{Ducloue:2019ezk}. The Balitsky-Kovchegov evolution equation (BK)~\cite{Balitsky:1995ub,Kovchegov:1999yj} has been used with great success to describe the internal structure and dynamics of protons in the impact parameter independent framework~\cite{Albacete:2010sy}. This evolution equation can be schematically written as $\partial_y N=K\otimes (N-N^2)$. It incorporates non-linear dynamics via the second term proportional to $N^2$. Omitting this term, the BK equation becomes equivalent to BFKL equation, which has been shown to give a satisfactory description of HERA data \cite{Avsar:2009pv}. 

The solution of the BK equation --- the dipole scattering amplitude $N$ --- allows us to calculate a wide spectrum of observables e.g.~\cite{Albacete:2010sy,vanHameren:2016ftb}. In our previous work, we have lifted one of the common approximations that were needed for solving this equation and by utilizing the collinearly improved kernel, we have shown that the impact parameter dependent computation can be obtained without being spoiled by the non-perturbative effects of Coulomb tails~\cite{Bendova:2019psy}.
In this work, we focus on quantitatively addressing the onset of saturation effects in nuclear targets by suppressing the non-linear term in the equation using our recently proposed initial condition. We are aware that the applicability of our model is restricted to sufficiently high energies where gluons are dominant scattering targets but not in the asymptotic limit where Coulomb tails would reappear~\cite{Contreras:2019vox}. This is however outside of reach of current and future planned facilities and thus the resulting signals are of interest for the physics program planned such experiments.

\section{Balitsky-Kovchegov equation}
The leading order Balitsky-Kovchegov evolution equation \cite{Balitsky:1995ub,Kovchegov:1999yj} for the impact parameter dependent case with the assumption of identical scattering amplitude for various angles between the transverse dipole size vector $\vec{r}$ and impact parameter vector $\vec{b}$ can be written as
\begin{eqnarray}
& & \frac{\partial N(r,b;y)}{\partial y} = \int d\vec{r_{1}}K(r,r_{1},r_{2})\big(N(r_{1},b_1;y)+  
\nonumber \\
&&N(r_{2},b_2;y) - N(r,b;y) - N(r_{1},b_1;y)N(r_{2},b_2;y)\big).
\label{fullbalitsky}
\end{eqnarray}
The evolution runs in rapidity $y=\log(x_0/x)$, where $x$ is the Bjorken variable and $x_0$ gives the initial value of Bjorken variable for the evolution. In order to solve the BK equation with explicit impact parameter dependence and to avoid the unphysical growth of so-called Coulomb tails \cite{Cepila:2018faq} originating from the non-perturbative region of its phase space we shall use the collinearly improved kernel~\cite{Iancu:2015joa} expressed as
\begin{equation}\label{collinearlyimproved}
K(r, r_1, r_2)  = \frac{\overline{\alpha}_s}{2\pi}\frac{r^{2}}{r_{1}^{2}r_{2}^{2}} \left[\frac{r^{2}}{\min(r_{1}^{2}, r_{2}^{2})}\right]^{\pm \overline{\alpha}_sA_1} \frac{J_1(2\sqrt{\overline{\alpha}_s \rho^2})}{\sqrt{\overline{\alpha}_s \rho}}.
\end{equation}
with the smallest-dipole prescription for its running coupling
\begin{equation}\label{alph}
\alpha_{s} (r) = \frac{4\pi}{\beta_{0,n_{f}}\ln\left(\frac{4C^{2}}{r^{2}\Lambda ^{2}_{n_{f}}}\right)},
\end{equation}
as described in~\cite{Bendova:2019psy} with all the parameter values and in greater detail.\\
In order to solve the BK equation for the nuclear case, one has to start with a nuclear initial condition. We have chosen to treat individually the dependence on the transverse size of the dipole $\vec{r}$ and the dependence on the distance of the dipole from the center of the target $\vec{b}$. For the $r$-dependence, we have parametrized our initial condition as in the GBW model~\cite{GolecBiernat:1998js} and for the $b$-dependence, we have chosen to mimic the density profile of the target parametrized by the Woods-Saxon distribution expressed as \cite{Cepila:2020xol}
\begin{equation}
\rho_A(b,z) = \rho_0\frac{1}{\exp\left[(r-{\rm R})/{\rm a}\right]+1},
\label{eq:WS}
\end{equation}
where $r\equiv\sqrt{b^2+z^2}$ and parameters are given by \cite{DeJager:1987qc}. In order to obtain the nuclear thickness of the target, one has to integrate the Woods-Saxon distribution over the longitudinal coordinate $z$ as
 \begin{equation}
 T_A(b) = \int\limits_{-\infty}^{+\infty} dz \rho_A(b,z).
 \end{equation}
Then we can define our nuclear initial condition (we denote this model as b-BK-A in the plots) as
\begin{equation}
\label{eq:nuclear_ic}
N^A(r, b,y=0) =  1 - \exp\left(-\frac{Q^2_{s0}(A)}{4}r^2 \frac{T_A(b_{q_1},b_{q_2})}{2}\right)
\end{equation}
with
\begin{equation}
T_A(b_{q_1},b_{q_2})= \frac{1}{T_A(0)} \left[T_A(b_{q_1})+T_A(b_{q_2})\right],
\end{equation}
where the term $1/T_A(0)$ normalizes the integrated distribution so that it reaches 1 at $b=0$. The values of $Q^{2}_{s0}(Ca)=0.341$\,GeV$^2$ and $Q^{2}_{s0}(Pb)=$ \\ 0.609\,GeV$^2$ were taken from \cite{Cepila:2020xol}. A similar approach as Eq.~\ref{eq:nuclear_ic} has been used in the past along with a model for $x$-dependence of $Q^2_{s0}$ to describe the DIS nuclear HERA data~\cite{Armesto:2002ny,Kowalski:2003hm}.

\section{Nuclear structure functions}
An observable that is often used to describe the onset and characteristic of nuclear effects is the so-called nuclear modification factor. This variable tells us how much a nucleus differs from a simple sum of the constituent nucleons and is obtained in our framework with the use of the structure function that can be expressed in the dipole model \cite{Mueller:1993rr,Mueller:1989st,Nikolaev:1990ja} as
\begin{equation}\label{F22}
\hspace*{-0.2cm} F^A_2(x,Q^2) = \frac{Q^{2}}{4\pi^{2}\alpha_{\rm em}}\hspace*{-0.07cm}\sum_{i}\hspace*{-0.1cm}\int\hspace*{-0.1cm} d\vec{r}dz |\Psi^{i}_{T,L}(z, r)|^{2} \sigma^A_{q\bar{q}}(r,\tilde{x}_i).
\end{equation}
Here $\tilde{x}_i = x (1 + (4m^2_{q_i})/Q^2)$ with $m_{q_{i}}$  the mass of the $i$-quark~\cite{GolecBiernat:1998js}. The cross section of the interaction of the color dipole with the target can be obtained due to the optical theorem as
\begin{equation}
\sigma^A_{q\bar{q}}(r,x) = 2\int\mathrm{d}\vec b N^A(r,b, x).
\label{dipole-cs}
\end{equation}
The wave function representing the probability of a virtual photon splitting into a quark-antiquark dipole can be written \cite{Nikolaev:1990ja} as
\begin{equation}
|\Psi_{T}^{i}(z, r)|^{2}=\frac{3\alpha_{\rm em}}{2\pi^{2}}e_{q_{i}}^{2}\big((z^{2} + (1-z)^{2})\epsilon^{2}K^{2}_{1}(\epsilon r) + m_{q_{i}}^{2}K^{2}_{0}(\epsilon r)\big)
\end{equation}
 and 
\begin{equation}
|\Psi_{L}^{i}(z, r)|^{2} = \frac{3\alpha_{\rm em}}{2\pi^{2}}e_{q_{i}}^{2}\Big(4Q^{2}z^{2}(1-z)^{2}K^{2}_{0}(\epsilon r)\Big)
\end{equation}
for the transverse and longitudinal polarization of the incoming photon, respectively, and
$|\Psi_{T,L}^{i}(z,r)|^{2}$ is a sum of squares of both contributions. $K_{0}$ and $K_{1}$ are the MacDonald functions, $z$ is the fraction of the total photon longitudinal momentum carried by the quark, $e_{q_i}$ is the fractional charge in units of elementary charge of quark $i$, $\alpha_{\rm em}$ = 1/137 and $\epsilon^{2} = z(1-z)Q^{2} + m_{q_{i}}^{2}$. The quark masses were set to 100\,MeV$/c^2$ for light, 1.3\,GeV$/c^2$ for charm,  and 4.5\,GeV$/c^2$ for bottom quark. After computing the structure function in such way and after taking the proton structure function calculated in a similar way (see \cite{Bendova:2019psy,Cepila:2018faq}), one can obtain the nuclear modification factor as
\begin{equation}\label{eq:rpa}
R_{pA}\equiv \frac{F^A_2(x,Q^2)}{A\;F^p_2(x,Q^2)}.
\end{equation}
The longitudinal structure function can be within the same model expressed as 
\begin{equation}\label{FL}
\hspace*{-0.2cm} F^A_L(x,Q^2) = \frac{Q^{2}}{4\pi^{2}\alpha_{\rm em}}\hspace*{-0.07cm}\sum_{i}\hspace*{-0.1cm}\int\hspace*{-0.1cm} d\vec{r}dz|\Psi^{i}_{L}(z, \vec{r})|^{2}\sigma^A_{q\bar{q}}(\vec{r},\tilde{x}_i).
\end{equation}

\section{Results}
We have solved the BK equation in the impact-parameter dependent, collinearly improved framework. We have done so with the same initial condition for two cases $i)$ with the inclusion of the saturation effects represented by the non-linear term in Equation~\ref{fullbalitsky} and $ii)$ without the nonlinear term in order to understand the expected role of saturation in the solutions of this equation. The initial condition was chosen so that it resembles the transverse profile of the nucleus.
\begin{figure}
\begin{center}
\resizebox{0.45\textwidth}{!}{\includegraphics{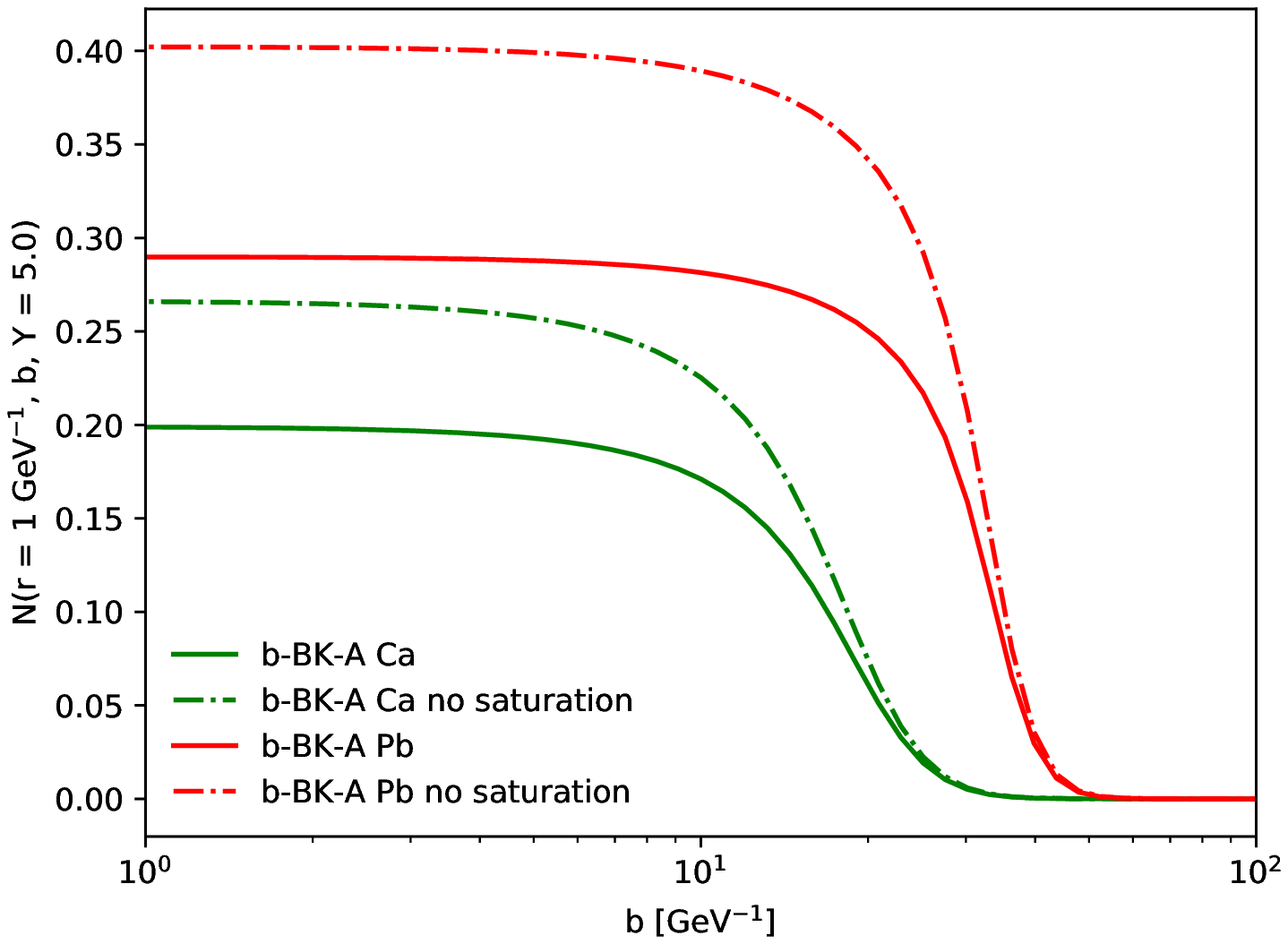}}
\resizebox{0.45\textwidth}{!}{\includegraphics{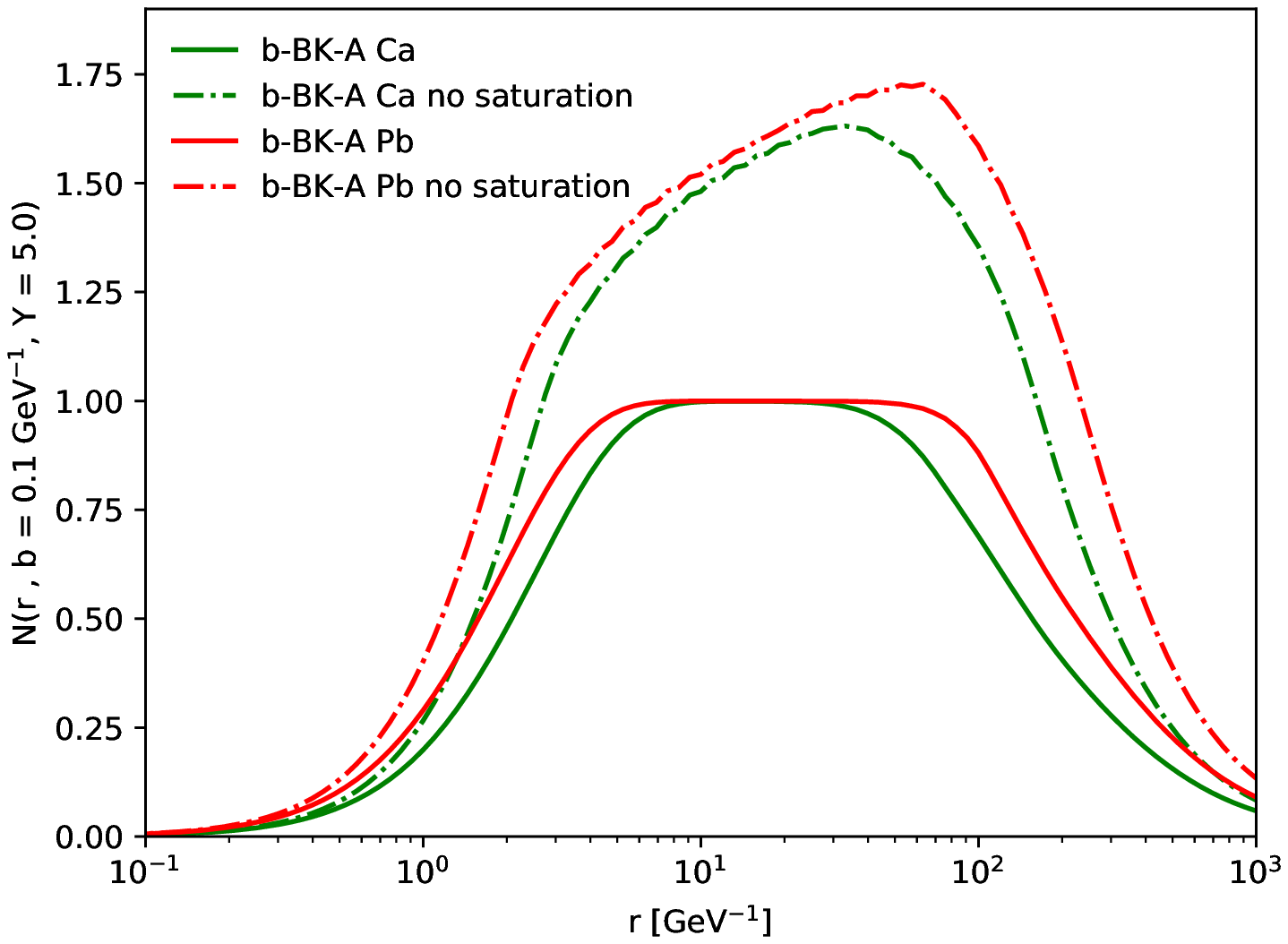}}
\caption{The dipole scattering amplitudes computed for Ca (green) and Pb (red) with (solid) and without (dashed) saturation effects. The comparison is done at $y=5$ as a function of the impact parameter for a dipole size $r=1$/GeV (upper) and as a function of the dipole size for an impact parameter $b=0.1$/GeV (lower).  \label{fig:ninb}}     
\end{center}
\end{figure}
Fig.~\ref{fig:ninb} shows the resulting scattering amplitude at $ y=5 $ for two nuclei (lead and calcium) and its dependence on the transverse dipole size $r$ for a fixed $b=0.1\,\mathrm{GeV}^{-1}$ and on the impact parameter $b$ for a fixed $r=1\,\mathrm{GeV}^{-1}$. We can see that the value of the non-saturated scattering amplitude exceeds unity. The difference between the non-linear and linear evolution is 30\%-60\%. 
\begin{figure}
\begin{center}
\resizebox{0.45\textwidth}{!}{\includegraphics{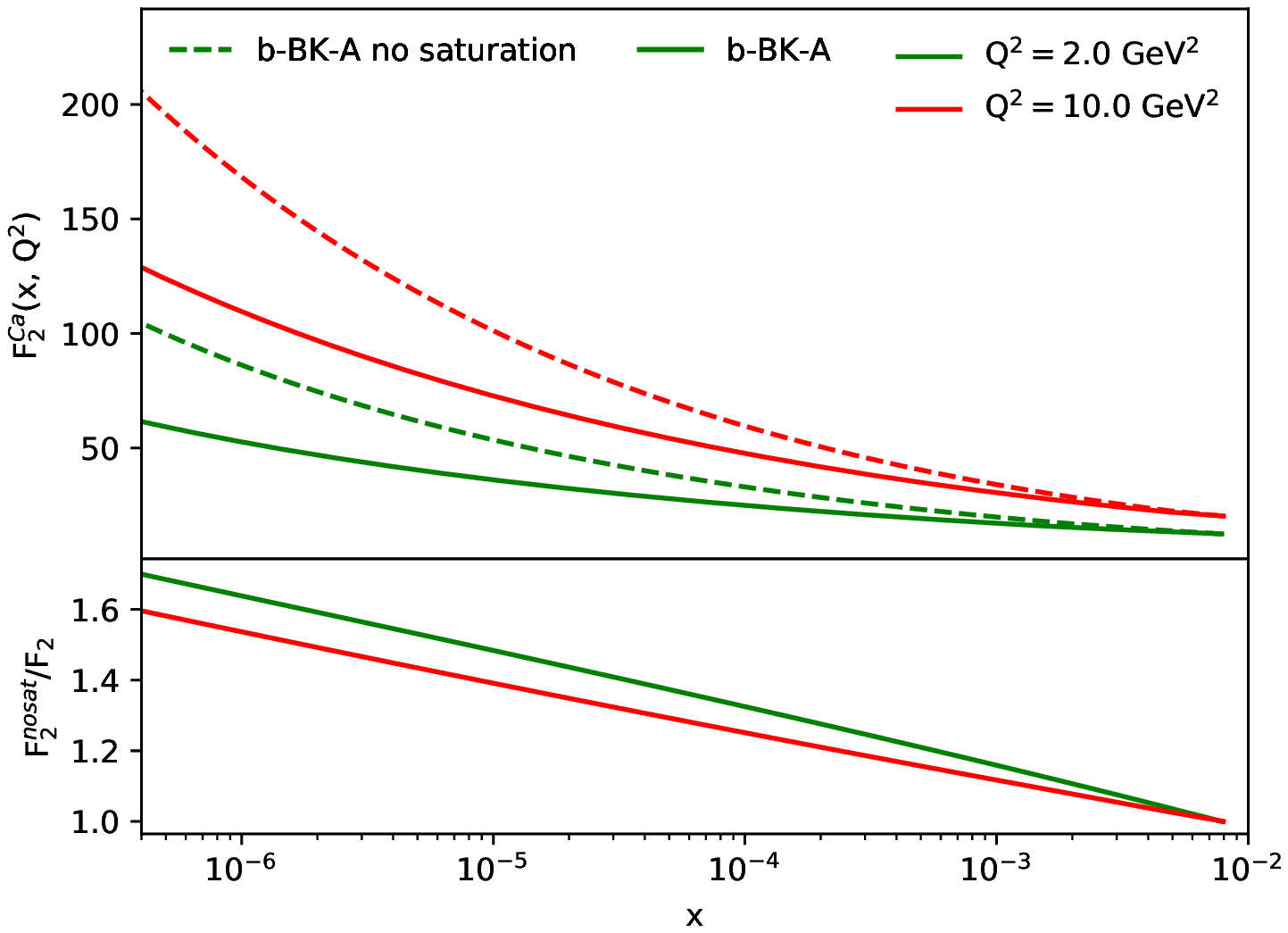}}
\resizebox{0.45\textwidth}{!}{\includegraphics{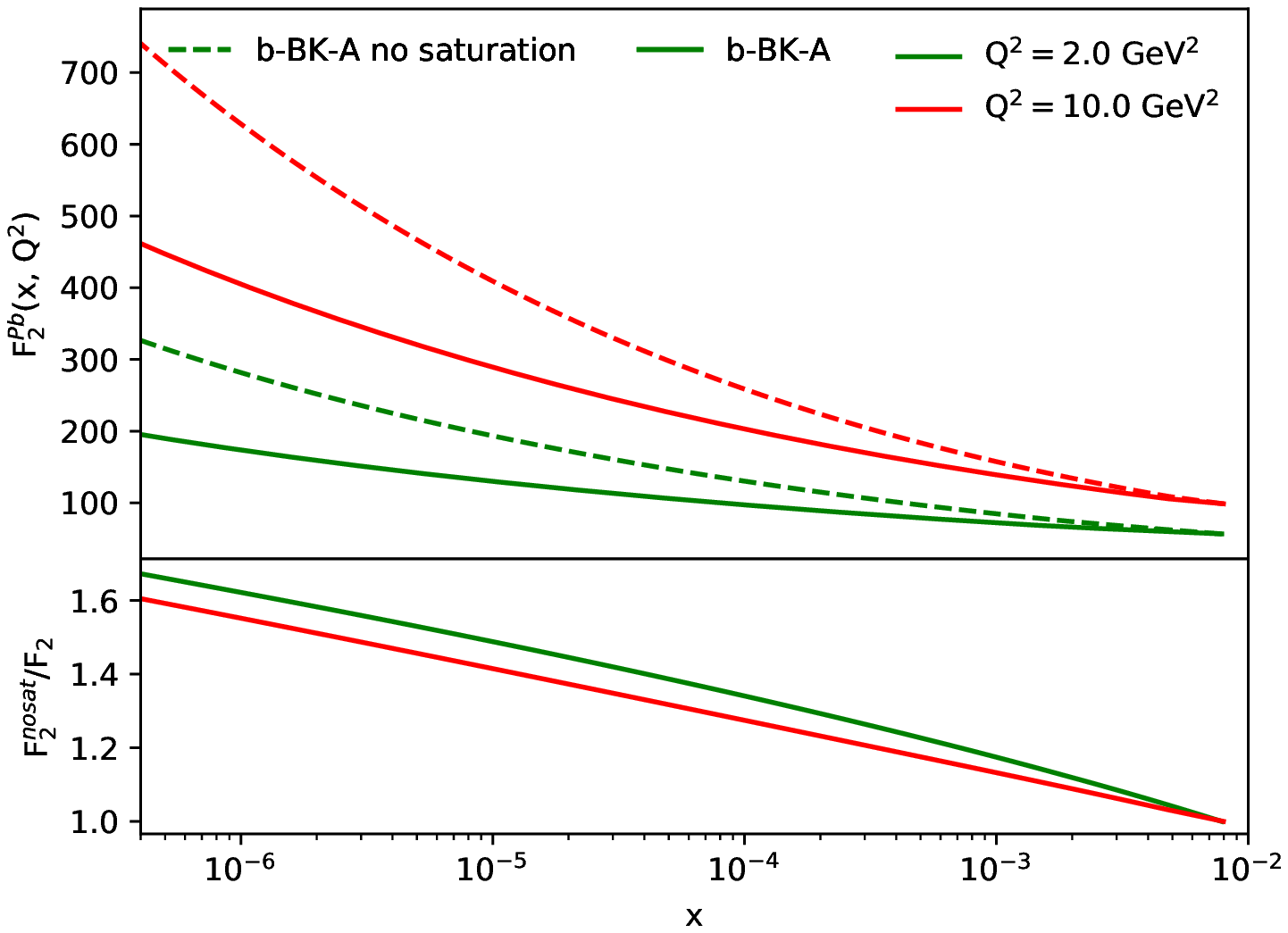}}
\caption{The nuclear structure function $F^A_2(x,Q^2)$ w.r.t $x$ computed with the Balitsky-Kovchegov evolution equation with and without saturation for two values of $Q^2$ for calcium (upper) and lead (lower). Bottom panel in the figures shows the ratio of the computation with and without saturation. \label{fig:F2Aca}}     
\end{center}
\end{figure}
\begin{figure}
\begin{center}
\resizebox{0.45\textwidth}{!}{\includegraphics{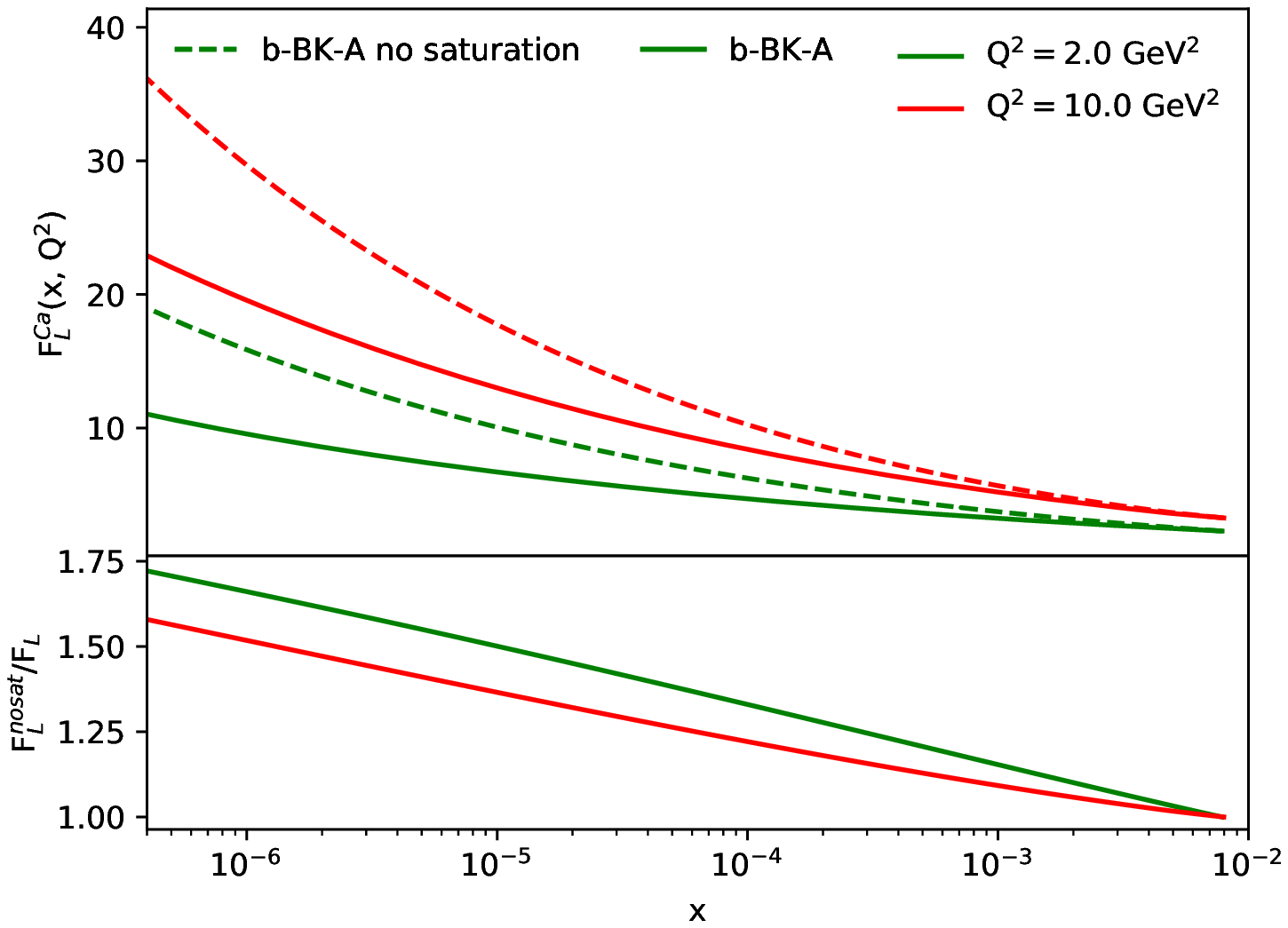}}
\resizebox{0.45\textwidth}{!}{\includegraphics{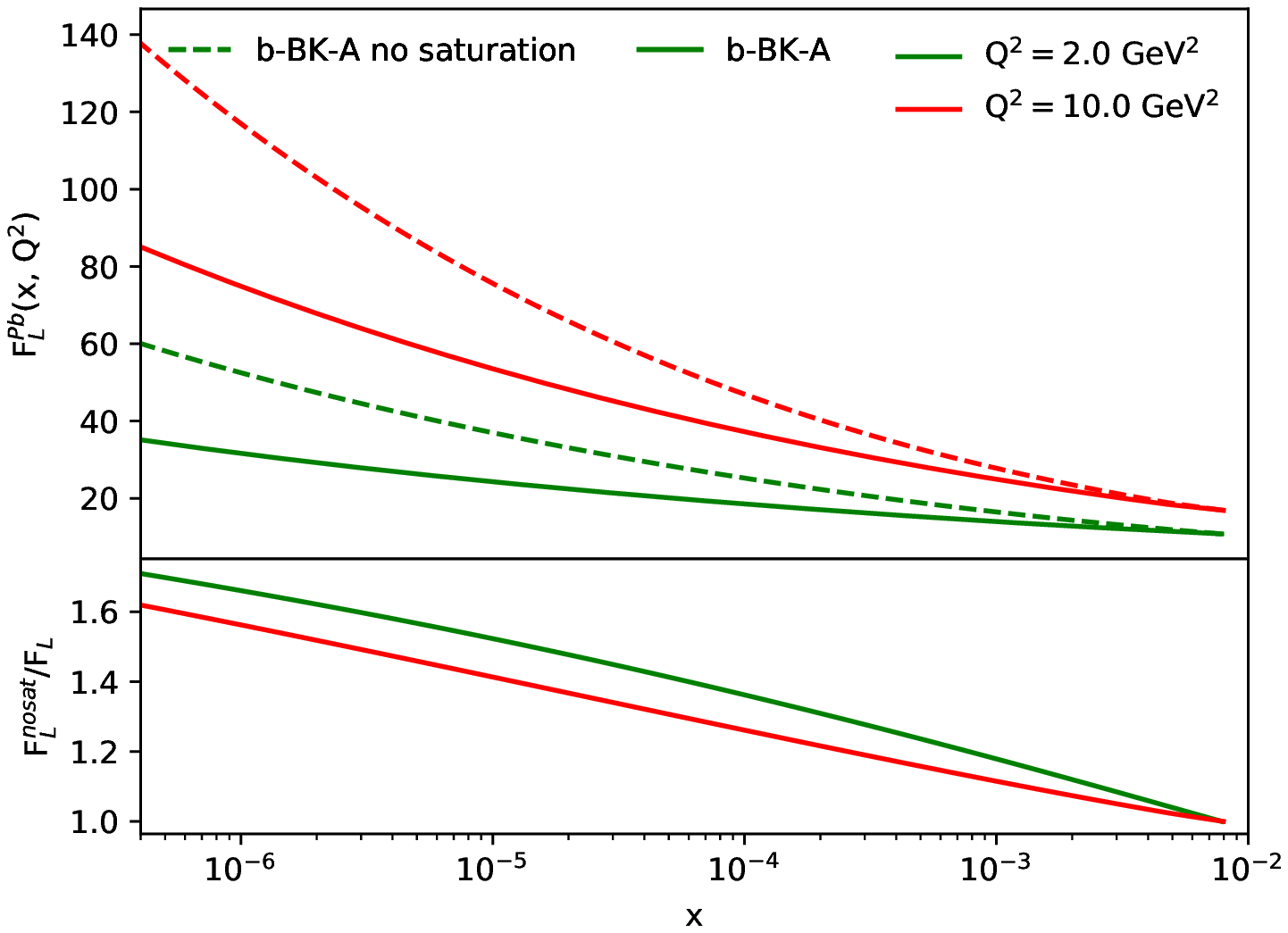}}
\caption{The nuclear structure function $F^A_L(x,Q^2)$ w.r.t $x$ computed with the Balitsky-Kovchegov evolution equation with and without saturation for two values of $Q^2$ for calcium (upper) and lead (lower). Bottom panel in the figures shows the ratio of the computation with and without saturation. \label{fig:Fl2Aca}}     
\end{center}
\end{figure}
In Figs.~\ref{fig:F2Aca} and \ref{fig:Fl2Aca} we show the computed structure functions $F_2(x, Q^2)$ and $F_L(x, Q^2)$ for calcium and lead respectively in linear and non-linear scenario as well as their ratios for two choices of $Q^2$. One can see that the non-linear evolution suppresses the structure functions and the difference grows with decreasing Bjorken $x$ reaching a value of about 40\% for $Q^2$ = 2\,GeV$^2$ at $x$ = $4\cdot10^{-5}$. At large Bjorken $x$ the difference is very small and thus one cannot discriminate between both scenarios using available data. Also, the difference decreases with increasing $Q^2$ both for $F_L(x, Q^2)$ and $F_2(x, Q^2)$. For $F_L(x, Q^2)$ the difference is greater than for $F_2(x, Q^2)$ at all scales and Bjorken $x$.
\begin{figure}
\begin{center}
\resizebox{0.45\textwidth}{!}{\includegraphics{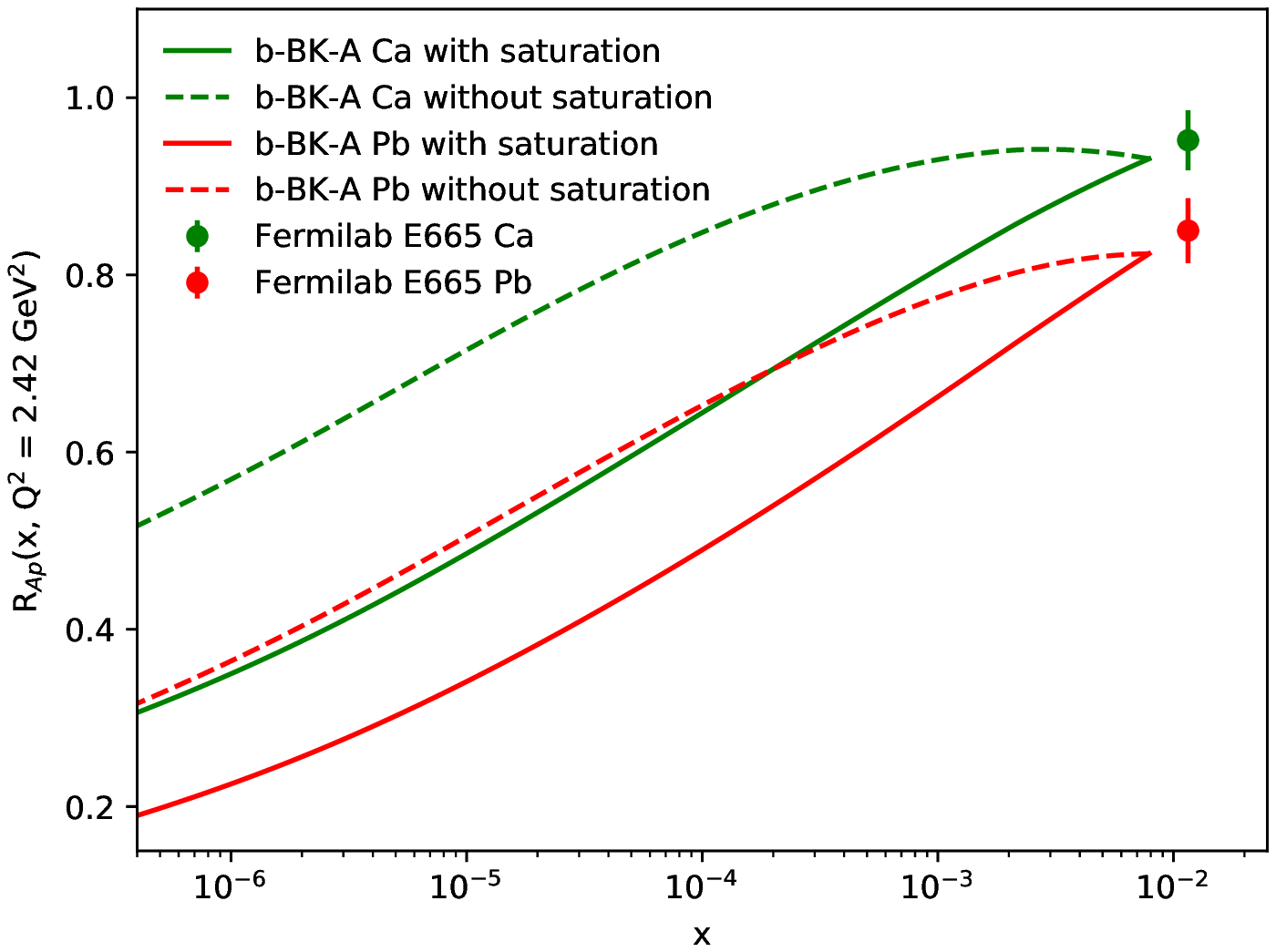}}
\resizebox{0.45\textwidth}{!}{\includegraphics{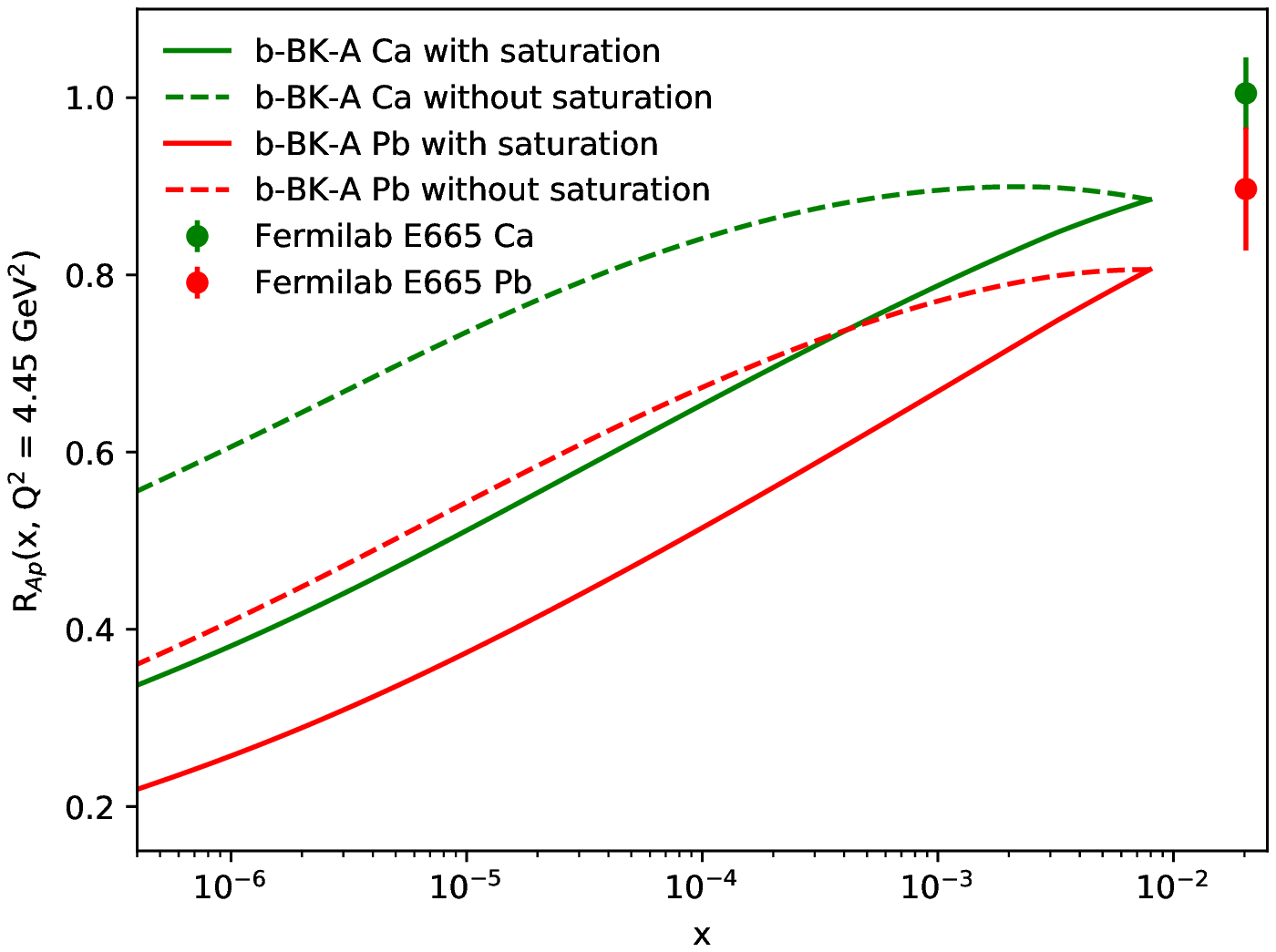}}
\caption{Nuclear modification factor computed for Ca (green) and Pb (red) with (solid) and without (dashed) saturation effects for $Q^2=2.42$ GeV$^2$ (upper) and $Q^2=4.45$ GeV$^2$ (lower). The predictions are compared with data from~\cite{Adams:1995is}.   \label{fig:rpa242}}     
\end{center}
\end{figure}
In Fig.~\ref{fig:rpa242}, we show the dependence of the nuclear modification factor on $x$ for calcium and lead obtained with the use of Eq.~(\ref{eq:rpa}) and compare it to data for $Q^2=2.42\,\mathrm{GeV}^2$ and $Q^2=4.45\,\mathrm{GeV}^2$. We can see, that the non-saturated scattering amplitudes produce larger nuclear modification factor implying softer nuclear effects. The difference between linear and non-linear model grows with decreasing Bjorken $x$ and so one can clearly discriminate between saturated and non-saturated model with future data from electron ion colliders. At large Bjorken $x$, both models are indistinguishable and both agree quite well with measured data point from E665 from Fermilab~\cite{Adams:1995is}.

\section{Conclusions}
In this paper we have presented a calculation of nuclear structure functions using the impact parameter dependent solution of the non-linear BK evolution equation. We have compared the resulting structure functions $F_2(x, Q^2)$ and $F_L(x, Q^2)$ and nuclear modification factor $R_{pA}(x,Q^2)$ with and without the non-linear term in BK evolution equation. The difference of the results with and without saturation is clearly visible and it rises with decreasing Bjorken $x$ and with scale $Q^2$ indicating that we will be able to distinguish between these two models with future data from electron-ion colliders.

\section{Acknowledgements}
This work has been supported from grant LTC17038 of the INTER-EXCELLENCE program at the Ministry of Education, Youth and Sports of the Czech Republic and the COST Action CA15213 THOR.
\bibliographystyle{epj}
\bibliography{biblio}
\end{document}